\input{aipcheck}

\documentclass[
	,numberedheadings % uncomment this option for numbered sections
  ]
  {aipproc}

\layoutstyle{6x9}

\begin{document}

\title{ULTRAVIOLET OBSERVATIONS OF SUPERNOVAE}


\author{Nino Panagia}{
  address={STScI, Baltimore, MD, USA;  panagia@stsci.edu\\ INAF -
  Observatory of Catania, Italy\\ Supernova Ltd., Virgin Gorda, BVI}
}

\begin{abstract}
The motivations to make ultraviolet (UV) studies of supernovae (SNe) are
reviewed and discussed in the light of the results obtained so far by
means of IUE and HST observations.  It appears that UV studies of SNe
can, and do lead to fundamental results not only for our understanding
of the SN phenomenon, such as the kinematics and the metallicity of the
ejecta, but also for exciting new findings in Cosmology, such as the
tantalizing evidence for "dark energy" that seems to pervade the
Universe and to dominate its energetics. The need for additional and
more detailed UV observations is also considered and discussed.

\end{abstract}

\maketitle

%%%%%%%%%%%%%%%%%%%%%%%%%%%%%%%%%%%%%%%%%%%%
%% MAINMATTER
%%%%%%%%%%%%%%%%%%%%%%%%%%%%%%%%%%%%%%%%%%%%

\section{Introduction} 

Supernovae (SNe) are the explosive death of massive stars as well as
moderate mass stars in binary systems.  They enrich the interstellar
medium of galaxies of most heavy elements (only C and N can efficiently
be produced and ejected into the ISM by red giants winds and by
planetary nebulae, as well as pre-SN massive star winds): nuclear
detonation supernovae, i.e., Type Ia SNe (SNIa), provide mostly Fe and
iron-peak elements, while core collapse supernovae, i.e., Type II (SNII)
and Type Ib/c (SNIb/c), mostly O and alpha-elements (see below for type
definitions). Therefore, they are the primary factors to determine the
chemical evolution of the Universe. Moreover, SN ejecta carry
approximately $10^{51}$ erg in the form of kinetic energy, which
constitute a large injection of energy into the ISM of a galaxy (for a
Milky Way class galaxy $E{^{MW}_{kin}}\simeq3 \times 10^{57}$ erg).  This
energy input is very important for the evolution of the entire galaxy,
both dynamically and for star-formation through cloud
compression/energetics.

In addition SNe are bright events that can be detected and studied up to
very large distances.  Therefore: (1) SN observations can be used trace
the evolution of the Universe. (2) SNe can be used as measuring sticks
to determine cosmologically interesting distances, either as "standard
candles" (SNIa, which at maximum are about 10 billion times bright than
the Sun, with a dispersion of the order of 10\%) or employing a refined
Baade-Wesselink method (SNII in which strong lines provide ideal
conditions for the application of the method, with a distance accuracy
of $\pm$20\%).  (3) Their intense radiation can be used to study the
ISM/IGM properties through measurements of the absorption lines.  Since
most of the strong absorption lines are found in the UV, this is best
done observing SNII at early phases, when the UV continuum is still
quite strong. Additional studies in the optical (mostly CaII and NaI
lines) are possible using {\bf all}  bright SNe.  However, only
combining optical and UV observations can one obtain the whole picture
and, therefore, SNII are the preferred targets for these studies. (4)
Finally, the strong light pulse provided by a SN explosion (the typical
HPW of a light curve in the optical is about a month for SNIa and about
two-three months for SNII; in the UV the light curve evolution is much
faster) can used to probe the intervening ISM in a SN parent galaxy by
observing the brightness, and the time evolution of associated light
echoes.

\section{Ultraviolet Observations}

The launch of the International Ultraviolet Explorer (IUE) satellite in
early 1978 marked the beginning of a new era for SN studies because of
its capability of measuring the ultraviolet emission from  objects as
faint as m$_B$=15.  Moreover, just around that time, other powerful
astronomical instruments became available, such as the Einstein
Observatory X-ray measurements, the VLA for radio observations, and a
number of telescopes either dedicated to infrared observations (e.g.
UKIRT and IRTF at Mauna Kea) or equipped with new and highly efficient
IR instrumentation (e.g. AAT and ESO observatories).  As a result,
starting in the late 70's a wealth of new information become available
that, thanks to the coordinated effort of astronomers operating at
widely different wavelengths, has provided us with fresh insights as for
the properties and the nature of supernovae of all types. Eventually,
the successful launch of the Hubble Space Telescope (HST) opened new
possibilities for the study of supernovae, allowing us to study SNe with
an accuracy unthinkable before and to reach the edge of the Universe. 

Even after 18 years of IUE observations and 16 more of HST observations,
the number of SN events that have been monitored with UV spectroscopy is
quite small and hardly include more than two objects per SN type and
hardly with good quality spectra for more than three epochs each. As a
consequence, we still know very little about the properties and the
evolution of the ultraviolet emission of SNe.  On the other hand, it is
just the UV spectrum of a SN, especially at early epochs, that contains
a wealth of valuable and crucial information that cannot be obtained
with any other means. Therefore, we truly want to monitor many more SNe
with much more frequent observations.  

We have learned at this Conference that SWIFT, in addition to doing UV
photometry, is also able to obtain low resolution spectra of SNe at
$\lambda>2000$Å with a sensitivity comparable to that of IUE and that
spectroscopic observations of SNe with SWIFT are in the planning (see,
e.g., the contribution by F. Bufano). This is an exciting possibility
that promises to provide very valuable results and to fill the gaps in
our knowledge about UV properties of SNe.

Here, I present a short summary of the UV observations of supernovae. A
more detailed review on this subject can be found in Panagia (2003).

\section{Type Ia Supernovae}

Type Ia supernovae are characterized by a lack of hydrogen in their
spectra at all epochs and by a number of typically broad, deep
absorption bands, most notably the Si II 6150Å  (actually the
blue-shifted absorption of the 6347-6371Å Si II doublet; see e.g.
Filippenko 1997), which dominate their spectral distributions at early
epochs.  SNIa are found in all types of galaxies, from giant ellipticals
to dwarf irregulars.  However, the SNIa explosion rate, normalized
relative to the galaxy H or K band luminosity and, therefore, relative
to the galaxy mass, is much higher, up to a factor of 16 when comparing
the extreme cases of irregulars and ellipticals (Della Valle \& Livio
1994, Panagia 2000, Mannucci et al. 2005) in late type galaxies than in
early type galaxies. This suggests that, contrary to common belief, a
considerable fraction of SNIa belong to a relatively young (age much
younger that 1$\sim$Gyr), moderately massive ($\sim$5M$_\odot$$<$ M(SNIa
progenitor)$<$ 8M$_\odot$) stellar population (Mannucci, Della Valle \&
Panagia 2006), and that in present day ellipticals SNIa are mostly the
result of capture of dwarf galaxies by massive ellipticals (Della Valle
\& Panagia, 2003, Della Valle et al. 2005)

%----------------------------------------------------------------
% figure1
\begin{figure}
\centerline{\includegraphics[width=9.5cm]{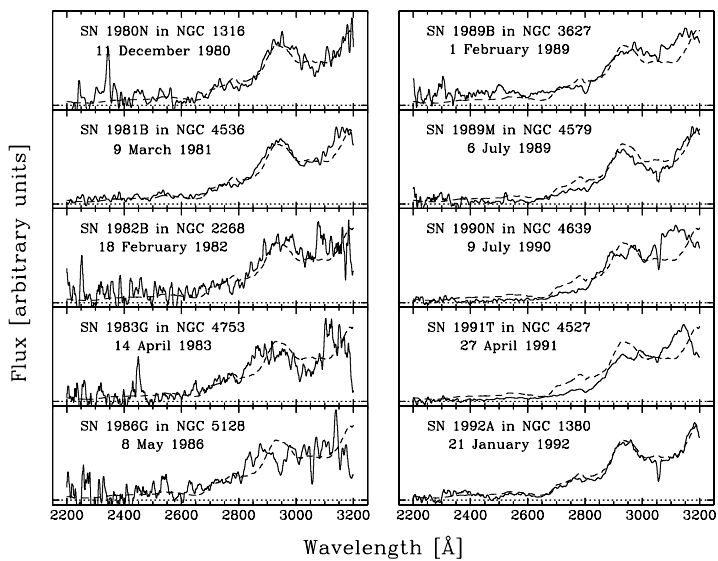} }
\caption{Ultraviolet spectra of ten Type Ia supernovae observed with IUE
around maximum light. The dashed line is SN~1992A spectrum as measured
with HST-FOS.
}
\end{figure}
%----------------------------------------------------------------

\subsection{Existing Samples of UV Spectra of SNIa}

Although 12 type Ia SNe were observed with IUE, only two events, namely
SN1990N and SN1992A, had extensive time coverage, whereas all others
were observed only around maximum light either because of their
intrinsic UV faintness or because of satellite pointing constraints.
Even so, one can reach important conclusions of general validity, which
are confirmed by the detailed data obtained for a few SNIa.

The UV spectra of type Ia SNe are found to decline rapidly with
frequency, making it hard to detect any signal at short wavelengths. 
This aspect is illustrated in Fig. 1, which displays the UV long
wavelength spectra of 10 type Ia SNe observed with IUE.  It appears that
the spectra do not have a smooth continuum but rather consist of a
number of  "bands'' that are observed with somewhat different strengths.
The fact that the spectrum is so similar for most of the SNe  supports
the idea of an overall homogeneity in the properties of type Ia SNe. 

On the other hand, some clear deviations from ``normal'' can be
recognized for some SNIa.  In particular, one can notice that both
SN1983G and SN1986G display excess flux around 2850 Å, and a deficient
flux around 2950 Å.  This suggests that the Mg II resonance line is much
weaker, which may indicate a lower abundance of Mg in these
fast-decline, under-luminous SNIa.  On the other hand, SN1990N, SN1991T,
and, possibly, SN1989M show excess flux around $\sim$2750 Å and
$\sim$2950 Å and a clear deficit around  $\sim$3100 Å, which may be
ascribed to enhanced Mg II and Fe II features in these slow-decline,
over-luminous SNIa.

The best studied SNIa event so far is the "normal'' type Ia supernova
SN1992A in the S0 galaxy NGC1380 that was observed as a TOO by both IUE
and HST (Kirshner et al. 1993). The HST-FOS spectra, from 5 to 45 days
past maximum light, are the best UV spectra available for a SNIa (see
Fig. 2) and reveal, with good signal to noise ratio, the spectral region
blueward of $\sim$2650 Å. 

%----------------------------------------------------------------
% figure2
\begin{figure}
\centerline{\includegraphics[width= 9cm]{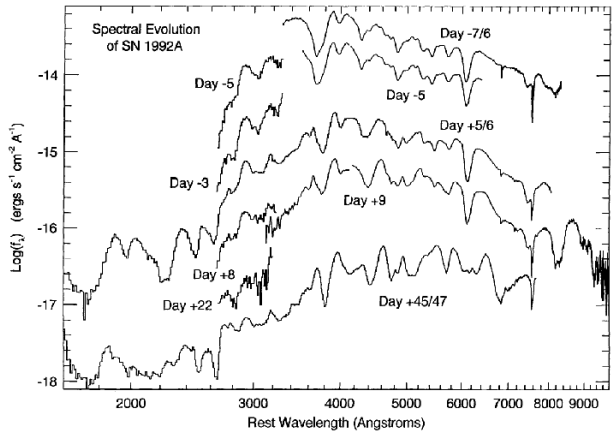} }
\caption{The spectral evolution  of SN1992A [adapted
from Kirshner et al. 1993].}
\end{figure}
%----------------------------------------------------------------

An LTE analysis of the SN1992A spectra shows that the features in the
region shortward of $\sim$2650Å are P Cygni absorptions due to blends of
iron peak element multiplets and the Mg II resonance multiplet. Newly
synthesized Mg, S, and Si probably extend to velocities at least as high
as $\sim$19,000 km~s$^{-1}$.  Newly synthesized Ni and Co may dominate the iron
peak elements out to  $\sim$13,000 km~s$^{-1}$ in the ejecta of SN1992A.  On
the other hand, an analysis of the O I $\lambda$7773 line in SN1992A and
other SNIa implies that the oxygen rich layer in typical SNIa extends
over a velocity range of at least $\sim$11,000-19,000 km~s$^{-1}$, but none of
the "canonical'' models has an O-rich layer that completely covers this
range. Even higher velocities were inferred by Jeffery et al. (1992) for
the overluminous, slow-decline SNIa SN1990N and SN1991T through an LT
analysis of their photospheric epoch optical and UV spectra.  In
particular, matter moving as fast as 40,000 and 20,000 km~s$^{-1}$ were found for 
SN1990N and SN1991T, respectively.

It thus appears that type Ia supernovae are consistently weak UV
emitters, and even at maximum light their UV spectra fall well below a
blackbody extrapolation of their optical spectra.  Broad features due to
P Cygni absorption of Mg II and Fe II are present in all SNIa spectra,
with remarkable constancy of properties for normal SNIa and systematic
deviations for slow-decline, over-luminous SNIa (enhanced Mg II and Fe
II absorptions) and fast-decline, under-luminous SNIa (weaker Mg II
lines).

\section{Core Collapse Supernovae: Types II and Ib/c}

Massive stars (M*$>$8M$_\odot$) are believed to end their evolution
collapsing over their inner Fe core and producing an explosion by a
gigantic bounce that launches a shock wave that propagates through the
star and eventually erupts through the progenitor photosphere, ejecting
several solar masses of material at velocities of several thousand
km~s$^{-1}$.  The current view is that single stars (as well as stars in wide
binary systems in which the companion does not affect the evolution of
the primary star) explode as type II supernovae, while supernovae of
types Ib and Ic originate from massive stars in interacting binary
systems.  Although the explosion mechanism is essentially the same in
both types, the spectral characteristics and light curve evolution are
markedly different among the different types.

\subsection{Type Ib/c Supernovae}

Type Ib/c supernovae (SNIb/c) are similar to SNIa in not displaying any
hydrogen lines in their spectra and are dominated by broad P Cygni-like
metal absorptions, but they lack the characteristic SiII 6150Å trough of
SNIa.  The finer distinction into SNIb and SNIc was introduced by
Wheeler and Harkness (1986) and is based on the strength of He I
absorption lines, most importantly He I 5876Å, so that the spectra of
SNIb display strong He I absorptions and those of SNIc do not.  SNIb and 
SNIc are found only in late type galaxies, often (but not always)
associated with spiral arms and/or H II regions.  They are generally
believed to be the result of the evolution of massive stars in close
binary systems. 

Although the properties of some peculiarly red and under-luminous SNI
(SN1962L and SN1964L) were already noticed by Bertola and collaborators
in the mid-1960s (Bertola 1964, Bertola et al. 1965), the first widely
recognized member and prototype of the SNIb class was SN1983N in
NGC5236=M83.  
 
%----------------------------------------------------------------
% figure3
\begin{figure}
\centerline{\includegraphics[width=10cm]{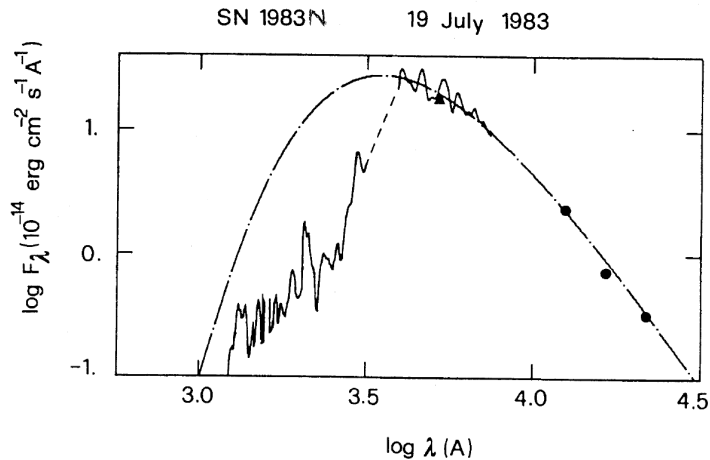} }
\caption{The spectrum of SN 1983N near maximum optical light, dereddened
with E(B-V)=0.16.  Both UV and optical spectra have been boxcar smoothed
with a 100Å~bandwith. The triangle is the IUE {\it Fine Error Sensor
(FES)} photometric point, and the dots represent the J, H, and K data.
The dash-dotted curve is a blackbody spectrum at T=8300K [adapted from 
Panagia 1985].
}
\end{figure}
%----------------------------------------------------------------

Because of its bright magnitude (B$\sim$11.6 mag at maximum light),
SN1983N is one of the best-studied SNe with IUE (see Panagia 1985). The
UV spectrum of SN1983N closely resembles that of type Ia SNe at
comparable epochs and, as such, only a minor fraction of the SN energy
is radiated in the UV. In particular, only $\sim$13\% of the total
luminosity was emitted by SN1983N shortward of 3400Å at the time of the
UV maximum.  Moreover, there is no indication of any stronger emission
in the UV at very early epochs; this implies that the initial radius of
the SN, i.e. the radius the stellar progenitor had when the shock front
reached the photosphere, was probably $<10^{12}$cm, ruling out a RSG
progenitor. From the bolometric light curve Panagia (1985) estimated
that $\sim$0.15 M$_\odot$ of  $^{56}$Ni was synthesized in the
explosion.

The best observed SNIc is SN1994I that was discovered on 2 April 1994
in the grand design spiral galaxy M51 and was promptly observed both
with IUE (as early as 3 April) and with HST- FOS (19 April).  The UV
spectra were remarkably similar to those obtained for SN1983N and,
although they were taken only at two epochs well past maximum light (10
days and 35 days), they were of high quality.  From synthetic spectra
matching the observed spectra from 4 days before to 26 days after the
time of maximum brightness, the inferred velocity at the photosphere
decreased from 17,500 to 7,000 km~s$^{-1}$ (Millard et al. 1999). Simple
estimates of the kinetic energy carried by the ejected mass gave values
that were near the canonical supernova energy of 10$^{51}$ erg. Such
velocities and kinetic energies for SN1994I are "normal'' for SNe and
are much lower than those found for the peculiar type Ic SN1997ef and
SN1998bw (see, e.g. Branch 2000) which appear to have been
hyper-energetic.

Thus, as type Ia, type Ib/c supernovae are weak UV emitters with their
UV spectra much fainter than a blackbody extrapolation of both optical
and NIR spectra, and their typical luminosity is about a factor of 4
lower than that of SNIa. The mass of  $^{56}$Ni synthesized in a typical
SNIb/c is, therefore, $\sim$0.15 M$_\odot$.

\subsection{Type II Supernovae}

Type II supernovae display prominent hydrogen lines in their spectra
(Balmer series in the optical) and their spectral energy distributions
are mostly a continuum with relatively few broad P Cygni-like lines
superimposed, rather than being dominated by discrete features as is the
case of all type I supernovae.  SNII are believed to be the result of a
core collapse of massive stars exploding at the end of their RSG phase. 
SN1987A was both a confirmation and an exception to this model.  It was
clearly the product of the collapse of a massive star, but it exploded
when it was a BSG, not an RSG. Since its properties are amply discussed
in many detailed papers presented at this Conference, we do not include
SN1987A in this summary of the UV properties of  "normal" SNII.

Among the other five SNII that were observed with IUE, only two, SN1979C and
SN1980K, were bright enough to allow a detailed study of their
properties in the UV (Panagia et al. 1980).  They were both of the
so-called "linear'' type (SNIIL), which is characterized by an almost
straight-line decay of the B and V-band light curves, rather than of the
more common "plateau'' type (SNIIP) which display a flattening in their
light curves starting a few weeks after maximum light.  

The SNII studied best in the UV so far is possibly SN1998S in NGC3877, a
type II with relatively narrow emission lines (SNIIn).  SN1998S was
discovered several days before maximum.  Its first UV spectrum, obtained
on 16 March 1998, near maximum light, was very blue and displayed lines
with extended blue wings, which indicate expansion velocities up to
18,000 km~s$^{-1}$ (Panagia 2003).  The UV spectral evolution of SN1998S (Fig.
5) showed the spectrum to gradually steepen in the UV, from near maximum
light on 16 March 1998 to about two weeks past maximum on 30 March, and
the blue absorptions to weaken or disappear completely.  About two
months after maximum (13 May 1998) the continuum was much weaker,
although its UV slope had not changed appreciably, and it had developed
broad emission lines, the most noticeable being the Mg II doublet at
about 2800Å.  This type of evolution is quite similar to that of SN1979C
(Panagia 2003) and suggests that the two sub-types are related
to each other, especially in their circumstellar interaction properties.

%----------------------------------------------------------------
% figure4
\begin{figure}
\centerline{\includegraphics[width=7.5cm]{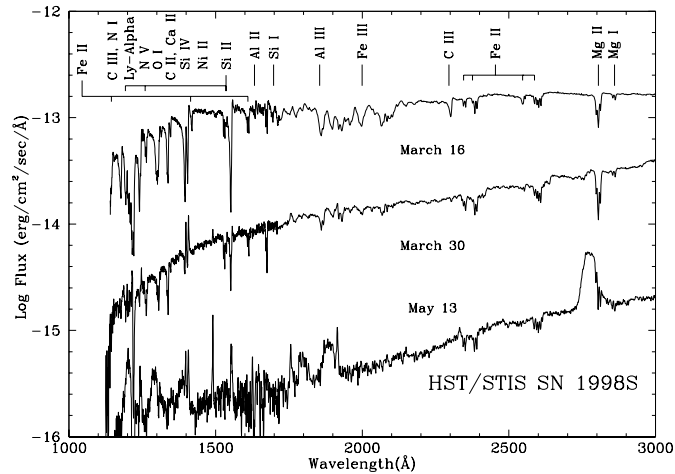} }
\caption{UV spectral evolution of SN1998S (SINS
project, unpublished). Shown are spectra obtained near maximum light
(March 16,1998), about two weeks past maximum (March 30, 1998), and
about two months after maximum (May 13, 1998).
}
\end{figure}
%----------------------------------------------------------------

A detailed analysis of early observations of SN1998S  (Lentz et al.
2001) indicated that early spectra originated primarily in the
circumstellar region itself, and later spectra are due primarily to the
supernova ejecta. Intermediate spectra are affected by both regions. A
mass-loss rate of order of $\sim10^{-4}$[v/(100km~s$^{-1}$)] M$_\odot$/yr was
inferred from these calculations but with a fairly large uncertainty.

Despite the fact that type II plateau (SNIIP) supernovae account for a
large fraction of all SNII, so far SN1999em in NGC1637 is the only SNIIP
that has been studied in some detail in the ultraviolet.  Although
caught at an early stage, SN1999em was already past maximum light (see,
e.g. Hamuy et al. 2001). An early analysis of the optical and UV spectra
(Baron et al. 2000) indicates that, spectroscopically, this supernova
appears to be a normal type II. Also, the analysis suggests the
presence of enhanced N as found in other SNII. 

Another sub-type of the SNII family is the so-called type IIb SNe,
dubbed so because at early phases their spectra display strong Balmer
lines, typical of type II SNe, but at more advanced phases the Balmer
lines weaken significantly or disappear altogether (see, e.g. Filippenko
et al. 1997) and their spectra become more similar to those of type Ib
SNe. A prototypical member of this class is SN1993J that was discovered
in early April 1993 in the nearby galaxy M81. An HST-FOS UV spectrum of
SN1993J was obtained on 15 April 1993, about 18 days after explosion,
and rather close to maximum light. The study of this spectrum (Jeffery
et al. 1994) shows that the approximately 1650-2900Å region is smoother
than observed for SN1987A and SN1992A and lacks strong P Cygni lines
absorptions caused by iron peak element lines. It is of interest to note
that the UV spectrum of SN1993J is appreciably fainter than observed in
most SNII, thus revealing its "hybrid'' nature and some resemblance to a
SNIb.  Synthetic spectra calculated using a parameterized LT procedure
and a simple model atmosphere do not fit the UV observations. Radio
observations suggest that SN1993J is embedded in a thick circumstellar
medium envelope (Van Dyk et al. 1994, Weiler et al. 2007). Interaction of
supernova ejecta with circumstellar matter may be the origin of the
smooth UV spectrum so that UV observations of supernovae could provide
insight into the circumstellar environment of the supernova progenitors.

Thus, despite their different characteristics in the detailed optical
and UV spectra, all type II supernovae of the various sub-types appear
to provide clear evidence for the presence of a dense CSM and, in many
cases, enhanced nitrogen abundance.  Their UV spectra at early phases
are very blue, possibly with strong UV excess relative to a blackbody
extrapolation of their optical spectra.

\section{Supernovae and Cosmology}

%----------------------------------------------------------------
% figure5
\begin{figure}
\centerline{\includegraphics[width=8.0cm]{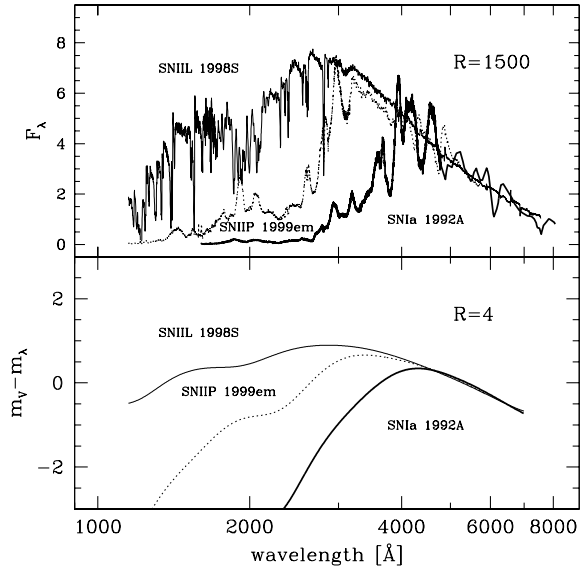} }
\caption{Top: Intermediate resolution (R=1500) spectra of three SNe near
maximum light, SNIa 1992A (Kirshner et al. 1993), SNIIn/L 1998S (Lentz
et al. 2001) and SNIIP 1999em (Baron et al. 2000) normalized in the V
band. Bottom: Low-resolution rendition of the observed spectra convolved
to a R=4 resolution showing the color differences in the UV. }
\end{figure}
%----------------------------------------------------------------

SNIa have gained additional prominence because of their cosmological
utility, in that one can use their observed light curve shape and color
to standardize their luminosities.  Thus, SNIa are virtually ideal
standard candles (e.g. Macchetto and Panagia 1999) to measure distances
of truly distant galaxies, currently up to redshift around 1 and,
considerably more in the foreseeable future. In particular, Hubble Space
Telescope observations of Cepheids in parent galaxies of SNIa (an
international project lead by Allan Sandage) have produced very accurate
determinations of their distances and the absolute magnitudes of normal
SNIa at maximum light that, in turn,  have lead to the most modern
measure of the  Hubble constant (i.e. the expansion rate of the local
Universe), H$_0=62.3\pm1.3(random)\pm5.0(systematic)$
km~s$^{-1}$Mpc$^{-1}$ (Sandage et al. 2006, and references therein).
This value is lower than the determination obtained by the   H$_0$
key-project from a combination of various methods, (H$_0=72\pm8$
km~s$^{-1}$Mpc$^{-1}$; Freedman  et al. 2001). The difference is 
well within the experimental uncertainties, and a weighted average of the
two determinations would provide a compromise value of H$_0=65.2\pm4.3$
km~s$^{-1}$Mpc$^{-1}$.

Observations of high redshift (i.e. z$>$0.1) SNIa have provided evidence
for a recent (past several billion years) acceleration of the expansion
of the Universe, pushed by some mysterious "dark energy". This is an
exciting result that, if confirmed, may shake the foundations of
physics.  The results of two competing teams (Perlmutter et al. 1998,
1999, Riess et al. 1998, Knop et al. 2003, Tonry et al. 2003, Riess et
al. 2004) appear to  agree in indicating a non-empty inflationary
Universe, which can be characterized by $\Omega_M\simeq0.3$ and
$\Omega_\Lambda\simeq0.7$. Correspondingly, the age of the Universe can
be bracketed within the interval 12.3-15.3 Gyrs to a 99.7\% confidence
level (Perlmutter et al. 1999).

However, the uncertainties, especially the systematic ones, are still
uncomfortably large and, therefore, the discovery and the accurate
measurement of more high-z SNIa are absolutely needed. This is a
challenging proposition, both for technical reasons, in that searching
for SNe at high redshifts one has to make observations in the near IR
(because of redshift) of increasingly faint objects (because of
distance) and for more subtle scientific reasons, i.e. one has to verify
that the discovered SNe are indeed SNIa and that these share the same
properties as their local Universe relatives.

One can discern Type I from Type II SNe on the basis of the overall
properties of their UV spectral distributions (Panagia 2003), because
Type II SNe are strong UV emitters, whereas all Type I SNe, irrespective
of whether they are Ia or Ib/c, have spectra steeply declining at high
frequencies (see Figure 5).   This technique of recognizing SNIa from
their steep UV spectral slope was devised by Panagia (2003), and has
been successfully employed by Riess et al. (2004a,b) to select their
best candidates for HST follow-up of high-z SNIa.   However, we have to
keep in mind  that by using this technique one is barely separatiung the
SNe with low UV emission (SNe Ia, Ib, Ic and, possibly, IIb) from the
ones with high UV emission (most type II SNe).  While it is a convenient
approach to select interesting candidates, it cannot be a substitute for
detailed spectroscopy, possibly at an R$>$100 resolution, to reliably
characterize the SN type.



\bibliographystyle{aipproc}   % if natbib is available
% \bibliographystyle{aipprocl} % if natbib is missing

\bibliography{sample}

%%%%%%%%%%%%%%%%%%%%%%%%%%%%%%%%%%%%%%%%%%%
%% The following lines show an example how to produce a bibliography
%% without the help of the BibTeX program. This could be used instead
%% of the above.
%%%%%%%%%%%%%%%%%%%%%%%%%%%%%%%%%%%%%%%%%%%

\end{document}